\documentclass[runningheads]{llncs}

\usepackage[T1]{fontenc}

\usepackage{graphicx}
\usepackage{algorithm}
\usepackage{algorithmic}
\usepackage{wrapfig,lipsum}
\usepackage{amsmath,bm}
\usepackage{url}            
\usepackage{booktabs}       
\usepackage{amssymb}
\usepackage{subfig}
\usepackage{cite}  
\usepackage{xcolor}
\usepackage{float}
\usepackage{amsmath}
\usepackage{multirow}
\usepackage{makecell}
\usepackage{color}
\usepackage{mathtools}
\usepackage{booktabs}
\usepackage{comment}
\usepackage{ifthen}
\usepackage{colortbl}
\usepackage{float} 
\usepackage{pifont}
\usepackage{enumitem}
\usepackage{threeparttable}
\usepackage{booktabs}
\usepackage{relsize}
\usepackage[T1]{fontenc}
\usepackage{wrapfig}
\usepackage{subfig}

\usepackage{orcidlink}
\usepackage[misc]{ifsym}

\usepackage{color}
\usepackage{hyperref}
\hypersetup{colorlinks}

\begin{document}

\title{Unsupervised Detection of Fetal Brain Anomalies using Denoising Diffusion Models}

\titlerunning{Unsupervised Detection of Fetal Brain Anomalies}

\authorrunning{M. D. S. Olsen et al.}

\author{Markus Ditlev Sjøgren Olsen\inst{1}\orcidlink{0009-0003-8896-9499}
\and
Jakob Ambsdorf\inst{2,4}\orcidlink{0000-0003-2925-8809}
\and
Manxi Lin\inst{1}\orcidlink{0000-0003-3399-8682}
\and
Caroline Taksøe-Vester\inst{3}\orcidlink{0000-0002-5634-2830}
\and
Morten Bo Søndergaard Svendsen\inst{1}\orcidlink{0000-0002-4492-3750}
\and
Anders Nymark Christensen\inst{1}\orcidlink{0000-0002-3668-3128}
\and
Mads Nielsen\inst{2,4}\orcidlink{0000-0003-1535-068X}
\and
Martin Grønnebæk Tolsgaard\inst{3}\orcidlink{0000-0001-9197-5564}
\and
Aasa Feragen\inst{1,4}\textsuperscript{(\Letter)}\orcidlink{0000-0002-9945-981X}
\and
Paraskevas Pegios\inst{1,4}\orcidlink{0009-0005-1471-4850}
}

\institute{
$^1$~Technical University of Denmark, Kongens Lyngby, Denmark\\
\email{\{afhar, ppar\}@dtu.dk}\\
$^2$~University of Copenhagen, Copenhagen, Denmark\\\
$^3$~CAMES, Rigshospitalet, Copenhagen, Denmark\\
$^4$~Pioneer Centre for AI, Copenhagen, Denmark}

\maketitle              

\begin{abstract}
Congenital malformations of the brain are among the most common fetal abnormalities that impact fetal development. 
Previous anomaly detection methods on ultrasound images are based on supervised learning, rely on manual annotations, and risk missing underrepresented categories. In this work, we frame fetal brain anomaly detection as an \emph{unsupervised} task using diffusion models. To this end, we employ an inpainting-based Noise Agnostic Anomaly Detection approach that identifies the abnormality using diffusion-reconstructed fetal brain images from multiple noise levels. Our approach only requires normal fetal brain ultrasound images for training, addressing the limited availability of abnormal data. Our experiments on a real-world \emph{clinical dataset} show the potential of using unsupervised methods for fetal brain anomaly detection. Additionally, we comprehensively evaluate how different noise types affect diffusion models in the fetal anomaly detection domain. 
\end{abstract}

\section{Introduction}
Congenital malformations of the brain are among the most common fetal developmental abnormalities, and their detection from ultrasound images is an important part of the mid-trimester fetal anomaly scan performed routinely around the world~\cite{paladini2007sonographic}. Detecting fetal brain anomalies using machine learning is challenging, as variations in image quality and probe position cause large variations in normal images~\cite{lin2024learning}, while abnormal images may differ only in small details~\cite{mishra2023dual}, giving poor separability of the two distributions. Further, the distribution of possible malformations is long-tailed, with many rare variations, and therefore little per-class training data.

\begin{figure}[t!]
    \centering
    \includegraphics[width=1\textwidth]{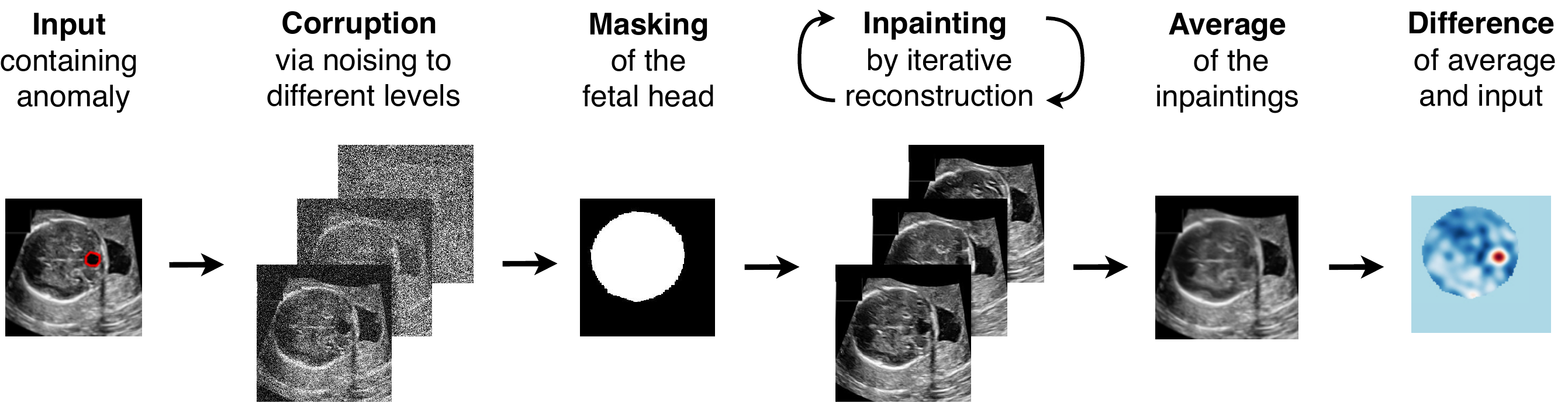}
    \caption{Overview of iNAAD for unsupervised detection of fetal brain anomalies. 
    }
    \label{fig:enter-label}
\end{figure}

Existing approaches~\cite{xie2020computer,xie2020using,lin2022use} have demonstrated the feasibility of \emph{supervised} detection of fetal brain anomalies. However, these methods (i) require labels for the individual malformations, sometimes down to anatomical details~\cite{lin2022use}, (ii) are bound to the detection of a closed set of frequent anomalies from the training data. To overcome these limitations, we present a proof-of-concept for the \emph{unsupervised} detection of fetal brain anomalies based on Denoising Diffusion Probabilistic Models~\cite{ho2020denoising} (DDPMs). Specifically, we adapt existing reconstruction-based methods~\cite{wyatt2022anoddpm,graham2023denoising} to build an inpainting-based Noise Agnostic Anomaly Detection (iNAAD) framework, involving averaging over reconstructions from multiple noise levels as in~\cite{graham2023denoising} and inpainting the fetal anatomy (see Fig.~\ref{fig:enter-label}). 
To the best of our knowledge, no prior work has investigated unsupervised detection methods for fetal brain anomalies. Our approach requires access only to ultrasound images of normal fetal brains during training, which are more readily available than abnormal cases.  In summary, we contribute 1) the first extensive evaluation of different noise types in DDPMs for the fetal ultrasound setting, 2) a diffusion-based algorithm iNAAD for unsupervised anomaly detection evaluated on a clinical dataset with a wide range of common fetal brain anomalies.

\section{Related Work}

Detecting developmental malformations from ultrasound images is a key goal of mid-trimester scans. Proposed methods include using biometry parameters from anatomical structures \cite{wu2020automatic,plotka2021fetalnet,sinclair2018human} or identifying expected normal structures \cite{komatsu2021detection} in fetal brains. The success of these methods, however, depends on auxiliary
detection models. Other approaches\cite{xie2020using,xie2020computer} focus on directly predicting abnormal brains using standard supervised binary classification methods. In \cite{lin2022use}, a multi-task framework is used to classify nine types of abnormalities and detect sub-features with bounding boxes. Yet, these methods are constrained to detecting only the most common malformations and require extensive data collection and preprocessing. In this work, we frame the task of fetal brain anomaly detection as an unsupervised problem by leveraging a large clinical dataset of normal fetal brain images without assuming prior knowledge of specific anomaly types.

Detecting fetal brain anomalies can be approached as an out-of-distribution (OOD) task, utilizing only in-distribution (ID) images of normal anatomy during training \cite{yang2021generalized}. Such methods, however, come with challenges of their own. Likelihood-based methods are prone to miscalibration~\cite{ren2019likelihood, nalisnick2019detecting} and adversarial attacks~\cite{fort2022adversarial}. Reconstruction-based methods, including VAE-based\cite{chen2018unsupervised}, compare inputs to their reconstructions, assuming more accurate results for ID samples. The success of DDPMs~\cite{ho2020denoising} opened up new opportunities in medical anomaly detection, 
by tailoring noise types\cite{wyatt2022anoddpm,frotscher2023unsupervised,kascenas2023role}
or using classifier guidance~\cite{dhariwal2021diffusion} in weakly supervised methods\cite{wolleb2022diffusion,li2023fast}. In fetal ultrasound, DDPMs have been successfully used for fetal brain image generation \cite{iskandar2023towards}, and counterfactual explanations \cite{pegios2024diffusion}. 
In \cite{mishra2023dual}, a dual-conditional DDPM that requires ID subclass information of different heart views both during training and inference is proposed for OOD detection of other anatomies from ID heart views in ultrasound videos.
In our work, we present a multi-reconstruction algorithm using unconditional DDPMs~\cite{ho2020denoising} for unsupervised OOD detection of fetal brain anomalies based on \cite{graham2023denoising}, integrating an inpainting step \cite{lugmayr2022repaint} to limit reconstruction changes in fetal brain and extensively evaluating different noise types \cite{kascenas2023role} for the fetal ultrasound setting.

\section{Method}

\subsection{Learning Distribution of Normal Brain Images with DDPMs}
We model the distribution of ID brain images $\mathcal{P}_{ID}$ using DDPMs~\cite{ho2020denoising}, enabling the generation and reconstruction of normal brain images. DDPMs consist of two processes: In the forward process, the image distribution is converted into a pre-defined noise distribution by adding noise $\epsilon\sim\mathcal{P}$ over $T$ steps. while in the reverse process, images can be generated by progressively denoising them. 

Formally, given a noise scheduler $\beta_t$ which controls the magnitude of noise added at step $t$, $\alpha_{t} := 1 - \beta_{t}$ and $\bar{\alpha}_{t} := \prod_{s=1}^{t} \alpha_{s}$, the forward process is defined,

\begin{equation}
   x_t=x _ { 0 } \sqrt { \bar { \alpha } _ { t } } + \sqrt { 1 - \bar { \alpha } _ { t } } \epsilon, \epsilon\sim\mathcal{P}
   \label{eq:forwardProcess}
\end{equation}
where $\epsilon$ represents noise from a pre-defined distribution $\mathcal{P}$ and $0 \leq t \leq T$ denotes the level of noise degradation. When $t$ is low, a significant amount of information from the original image is retained. No information is assumed to remain at $t=T$ and $x_T$ appears similar to pure noise. The reverse process consists of a Markov chain that iteratively removes noise using a denoiser $\epsilon_\theta(x_t,t)$,

\begin{equation}
 x _ { t - 1 } = \frac { 1 } { \sqrt { \alpha _ { t } } } \left( x _ { t } - \frac { 1 - \alpha _ { t } } { \sqrt { 1 - \bar { \alpha } _ { t } } } \epsilon _ { \theta } \left( x _ { t } , t \right) \right) + \beta _ { t } \epsilon, \epsilon\sim\mathcal{P}
 \label{eq:reverseProcess}
\end{equation}

We train a neural network $\epsilon_\theta$ to estimate the noise for a given image $x_t$ and then compare it to the actual noise $\epsilon\sim\mathcal{P}$ with the following objective,
\begin{equation}
\theta ^ { * } = \underset { \theta } { \arg \min } \mathbb { E } _ { x_0\sim \mathcal{P}_{ID},t\sim\mathcal{U}(0,T) }  \left\| \epsilon - \epsilon _ { \theta } \left( x_t, t \right) \right\| ^ { 2 } 
\label{eq:TrainingObj}
\end{equation}
where $x_t$ follows the forward process in Eq. (\ref{eq:forwardProcess}) and $\theta$ are learnable parameters.

In practice, $\mathcal{P}$ is typically a Gaussian distribution. However, recent studies \cite{wyatt2022anoddpm,frotscher2023unsupervised,kascenas2023role,naval2024disyre} have shown that alternative noise distributions can significantly impact and improve medical anomaly detection tasks. In this paper, we assess the effect of three distinct noise distributions, namely Gaussian~\cite{ho2020denoising}, Simplex~\cite{wyatt2022anoddpm}, and Pyramid~\cite{frotscher2023unsupervised}, on denoising diffusion models for fetal brain anomaly detection.

\subsection{iNAAD: Inpainting-based Noise Agnostic Anomaly
Detection}
Following~\cite{wolleb2022diffusion,wyatt2022anoddpm,graham2023denoising}, we adopt a reconstruction-based anomaly detection approach, aiming to reconstruct input images $x_0$ using DPPMs trained on normal, anomaly-free, fetal ultrasound scans. Specifically, we apply the forward process to corrupt $x_0$ to $x_{s}$, for a fixed $1 \leq s \leq T$, and then retrieve the reconstructed image from $x_{s}$ by the reverse process. Hyperparameter $s$  controls the level of noise degradation. Given the image $x_{t-1}$ at step $t-1$ in the forward process, we denote its corresponding reconstruction with the same steps in the reverse process as $\bar{x}_{t-1}$. The altered content between the input image and its reconstruction can therefore be interpreted as an anomaly indicator. To quantify these anomalies, we present the iNAAD algorithm, which is outlined in Alg.~\ref{alg:simple_algoritm_for_reconstruction}. 

Inspired by~\cite{lugmayr2022repaint}, we constrain the reconstruction within the region of interest, i.e., the fetal brain in the image with inpainting. In particular, we apply a binary mask $m$ obtained with a pre-trained segmentation model~\cite{lin2023dtu} to ignore all the variations beyond the fetal brain. 
Given a pre-defined noise distribution $\mathcal{P}$ and a trained denoiser $\epsilon _ { \theta }$, we define the inpainted reconstruction $\hat{x}_{t-1}$ by,

\begin{equation}
 \label{eq:inpaint}
 \centering 
 \begin{array} { l } x _ { t - 1 } = x _ { 0 } \sqrt { \bar { \alpha } _ { t } } + \sqrt { 1 - \bar { \alpha } _ { t } } \epsilon \\ \bar{x} _ { t - 1 } = \frac { 1 } { \sqrt { \alpha _ { t } } } \left( x _ { t } - \frac { 1 - \alpha _ { t } } { \sqrt { 1 - \bar { \alpha } _ { t } } } \epsilon _ { \theta } \left( x _ { t } , t \right) \right) + \beta _ { t } \epsilon \\ \hat{x} _ { t - 1 } = m \odot x _ { t - 1 } + ( 1 - m ) \odot \bar{x} _ { t - 1 }\end{array}
\end{equation}

During the forward process, the information content of $ x_0$ is controlled by the noise level $s$. Similar to~\cite{graham2023denoising}, we aggregate reconstructions obtained by degrading $x_0$ with a range of multiple noise levels $s \in S$. By reconstructing all corrupted versions of $x_0$ and averaging these reconstructions, we obtain a final reconstructed image $\bar{x}$ that integrates information from all reconstructed versions while reducing noise from individual reverse processes~\cite{graham2023denoising}.

Finally, for detecting abnormalities, the choice of the similarity metric between $x_0$ and $\bar{x}$ is essential. We observed that the similarity metrics such as LPIPS, used in~\cite{graham2023denoising}, were not effective for distinguishing abnormal from normal fetal images. Despite exploring other semantic similarity metrics~\cite{czolbe2023semantic} we empirically chose to utilize the standard pixel-based Structural Similarity Index (SSIM) which proved more effective for our task. 

iNAAD requires only normal fetal brain ultrasound images for training. It identifies abnormalities by aggregating diffusion-reconstructed fetal brain images from various noise levels, incorporating an inpainting step to limit reconstruction changes in the fetal brain. The proposed method is summarised in Alg.~\ref{alg:simple_algoritm_for_reconstruction}.

\begin{algorithm}
\small
\caption{iNAAD for unsupervised fetal brain anomaly detection.}
\begin{algorithmic}
\STATE \textbf{Input:} original $x_{0}$, binary mask $m$, noise distribution $\mathcal{P}$, model $\epsilon _ { \theta }$, noise levels $S$ 
\STATE \textbf{Output:} average reconstructed image $\bar{x}$, similarity metric between $x_{0}$ and $\bar{x}$
\FOR{$s$ in S}
\STATE Define time step $t := s$
\STATE Corrupt original image $x_{0}$ up to noise level $t$ by sampling from $\mathcal{P}$ (Eq. \ref{eq:forwardProcess})
\FOR{$t$ to 1}
\STATE Get inpainted reconstruction $\hat{x} _ {t - 1}$ using mask $m$ and model $\epsilon _ { \theta }$ (Eq. \ref{eq:inpaint})
\ENDFOR
\ENDFOR
\STATE \textbf{return} $\bar{x}=\frac{1}{|S|}\sum_{s\in\{S\}}\hat{x}_{0,s}$ and \textsl{similarity\_metric}($x_{0}, \bar{x}$) 
\end{algorithmic}
\label{alg:simple_algoritm_for_reconstruction}
\end{algorithm}

\section{Experiments and Results}

\subsubsection{Dataset.} 
We constructed our dataset using a pre-trained standard plane classifier~\cite{Manxis_model} to extract images from the Danish national fetal ultrasound screening database. This includes a large set of ID images for developing DDPMs and OOD images for validation and testing. For the ID images, we sampled 221,177 mid-trimester images from unique patients, identifying 14,268 brain images. From 43,297 images with central nervous system malformations, we identified 3557 brain images and randomly sampled one per patient, resulting in 492 OOD images. Finally, we divided a split of 13568/250/250 ID images for train/validation/test, keeping 200 for external ID testing, and a split of 250/242 of OOD images for validation/test.

\subsubsection{Models and implementation.}
We implement and evaluate the effect of three noise distributions in the fetal ultrasound setting: Gaussian~\cite{ho2020denoising}, Pyramid~\cite{frotscher2023unsupervised}, and Simplex~\cite{wyatt2022anoddpm}. These distributions range from least (Gaussian) to most correlated (Simplex), with the latter designed to enable multi-scale image reconstruction by varying perturbations across different regions. Following original implementations, we define Gaussian as $\epsilon \sim \mathcal{N}(\mu=0,\sigma^2=1)$, Pyramid as $\epsilon \sim \sum _ { i = 1 } ^ { 10 } 0.8^{ i } \cdot U \left( \epsilon ^ { ( i ) } ; H , W \right)$, where $U$ is a bilinear operator that upscales the image to dimensions $H \times W$, $\epsilon^{i}$ represents Gaussian noise with dimensions $h_i \times w_i$,
and 0.8 being the scaling factor, and Simplex $\epsilon \sim \operatorname { Simplex } \left( \nu = 2 ^ { - 6 } , N = 6 , \gamma = 0.8 \right)$ where $\nu$ is the starting frequency of noise regions, $N$ is the number of layers of noise with different frequency, and $\gamma$ is the decay of noise throughout the layers of noise. 
A DDPM is trained for each noise type using the ID training set, following the same model architecture and hyperparameters as in~\cite{pegios2024diffusion}, using 500 diffusion steps, and training for 200K iterations with batch size 20. Following~\cite{frotscher2023unsupervised}, during reconstructions with Pyramid noise, we corrupt images with Gaussian noise to better allow the model to remove anomalous image features.

\begin{figure}[b]
    \centering
     \includegraphics[width=0.8\linewidth]{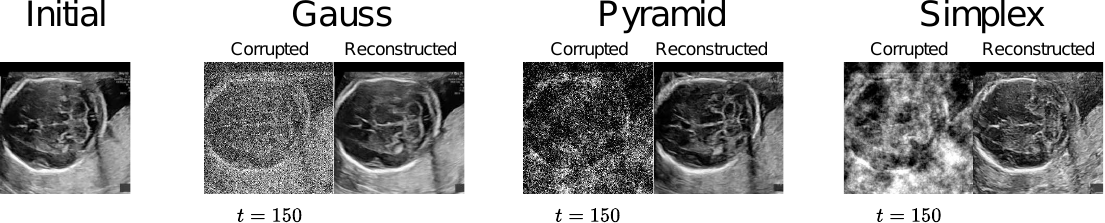}
      \caption{Reconstruction of a normal fetal brain from corruption level $t=150$.
  }
      \label{fig:nosie_forms}
  \end{figure}

\subsubsection{Evaluation of DDPMs.}
The ability to reconstruct ID images with high fidelity is an essential part of the approach, hence, we evaluate DDPMs both for generation and reconstruction. Table 1 compares DDPMs trained with different noise types for image generation based on FID using the external ID test and for image reconstruction across different corruption levels in terms of SSIM using the ID validation set. We observe that the reconstructive ability of DDPMs trained with Simplex and Pyramid decreases faster than Gaussian and they generate samples with lower fidelity. An example reconstruction is shown in Fig.~\ref{fig:nosie_forms}.

\begin{table}[t]
\centering
\caption{Evaluation for generation and reconstruction of normal fetal brains. }
\begin{tabular}{c|c|ccccccc}
\toprule

\multirow{2}{*}{Model} & \multirow{2}{*}{FID}  & \multicolumn{7}{c}{SSIM for different noise step levels $t$}\\

 & \multicolumn{1}{c|}{}& \multicolumn{1}{c}{50} & \multicolumn{1}{c}{75} & \multicolumn{1}{c}{100} & \multicolumn{1}{c}{150} & \multicolumn{1}{c}{200} & \multicolumn{1}{c}{250} & 300 \\
\midrule
DDPM-Gaussian &\textbf{48.39} & \multicolumn{1}{c|}{\textbf{0.989}}& \multicolumn{1}{c|}{\textbf{0.984}}& \multicolumn{1}{c|}{\textbf{0.979}}& \multicolumn{1}{c|}{\textbf{0.968}}& \multicolumn{1}{c|}{\textbf{0.955}}    & \multicolumn{1}{c|}{\textbf{0.933}}& \textbf{0.897} \\

DDPM-Pyramid &57.89& \multicolumn{1}{c|}{0.980}   & \multicolumn{1}{c|}{0.968}   & \multicolumn{1}{c|}{0.955}    & \multicolumn{1}{c|}{0.928}       & \multicolumn{1}{c|}{0.892}  & \multicolumn{1}{c|}{0.830}   &  0.752   \\
DDPM-Simplex &199.40 & \multicolumn{1}{c|}{0.981}   & \multicolumn{1}{c|}{0.967}   & \multicolumn{1}{c|}{0.948}    & \multicolumn{1}{c|}{0.905}    & \multicolumn{1}{c|}{0.836}    & \multicolumn{1}{c|}{0.753}    &    0.702 \\

\bottomrule 
\end{tabular}

\label{tab:fsim_ssim}
\end{table}

\subsubsection{Supervised baseline.}
A Resnet-18~\cite{he2016deep} architecture is used as a supervised baseline in the form of a binary classifier (normal/abnormal). We group all anomalies into one class due to the per-class scarcity. The model is initiated with ImageNet pre-trained weights and fine-tuned for 60 epochs using random augmentations during training, on the validation set (250 ID/250 OOD cases). We evaluate its performance on the final test set (250 ID/242 OOD cases).

\begin{table}[b]
    \centering
    \caption{AUROC results on the test set (250 ID/242 OOD cases) for iNAAD, and the Resnet-18 supervised baseline trained for binary classification, per anomaly group. The best scores are in bold, second best are underlined.}
  \resizebox{\linewidth}{!}{
    \begin{tabular}{c|c|c|c|c|c|c|c|c}
        \toprule
        \multicolumn{1}{c}{} & \multicolumn{8}{c}{ \textbf{AUROC per anomaly group}} \\ \midrule
         Model& \multicolumn{1}{c}{\thead{Microcephaly\\(n=40)}}&\multicolumn{1}{c}{\thead{Hydrocephalus\\(n=64)}}&\multicolumn{1}{c}{\thead{ACC\\(n=38)}}&\multicolumn{1}{c}{\thead{Cerebr. cyst\\(n=43)}}&\multicolumn{1}{c}{\thead{Ventriculomegaly\\(n=69)}}&\multicolumn{1}{c}{\thead{Spina bifida\\(n=39)}}&
         \multicolumn{1}{c}{\thead{Others\\(n=93)}}&\multicolumn{1}{c}{\thead{All\\(n=242)}}\\
         \midrule
         
 Resnet-18 &\textbf{0.63}&\textbf{0.69}&0.63&\underline{0.60}&\underline{0.71}&\textbf{0.78}&\textbf{0.65}&\textbf{0.67}\\ \midrule
 iNAAD-Gaussian &\underline{0.60}&0.62&\textbf{0.76}&0.53&\textbf{0.74}&0.57&\underline{0.60}&\underline{0.62}\\
 iNAAD-Simplex &0.56&\underline{0.65}&\underline{0.69}&\textbf{0.61}&0.69&\underline{0.62}&0.56&0.58\\
 iNAAD-Pyramid &0.60&0.64&0.68&0.57&0.69&\underline{0.62}&0.56&0.57\\
 \bottomrule

    \end{tabular}
    }
    \label{tab:ResultsCodes}
\end{table}

\begin{table}[t]
    \centering
    \caption{  Average Precision (AP) results on the test set (250 ID/242 OOD cases) for iNAAD, and the Resnet-18 supervised baseline trained for binary classification, per anomaly group. AP for random choice of normal cases is presented as Chance Level. The best scores are in bold, second best are underlined. 
    }
    \resizebox{\linewidth}{!}{
    \begin{tabular}{c|c|c|c|c|c|c|c|c}
    \toprule
        \multicolumn{1}{c}{} & \multicolumn{8}{c}{\textbf{AP per anomaly group}} \\ \midrule
         Models& \multicolumn{1}{c}{\thead{Microcephaly\\(n=40)}}&\multicolumn{1}{c}{\thead{Hydrocephalus\\(n=64)}}&\multicolumn{1}{c}{\thead{Acc\\(n=38)}}&\multicolumn{1}{c}{\thead{Cerebr. cyst\\(n=43)}}&\multicolumn{1}{c}{\thead{Ventriculomegaly\\(n=69)}}&\multicolumn{1}{c}{\thead{Spina bifida\\(n=39)}}&
         \multicolumn{1}{c}{\thead{Others\\(n=93)}}&\multicolumn{1}{c}{\thead{All\\(n=242)}}\\
         \midrule
  Chance Level & 0.10&0.17&0.9&0.11&0.18&0.09&0.19&0.48\\
        \midrule
  
  Resnet-18 &0.12&\textbf{0.37}&0.19&0.15&\textbf{0.42}&\textbf{0.30}&\textbf{0.29}&\textbf{0.65}\\\midrule
  iNAAD-Gaussian &0.17&0.28&\textbf{0.37}&0.11&\textbf{0.37}&0.17&\underline{0.28}&\underline{0.62}\\
  iNAAD-Simplex &\textbf{0.25}&\underline{0.33}&\underline{0.30}&\textbf{0.25}&0.36&\textbf{0.30}&0.27&0.56\\
  iNAAD-Pyramid &\underline{0.24}&0.32&\underline{0.30}&\underline{0.22}&0.36&0.27&0.26&0.55\\
 \bottomrule
    \end{tabular}
    }
    \label{tab:ResultsCodes2}
\end{table}

\subsubsection{Results.}
\begin{figure}[t]
    \centering
    \includegraphics[width=0.3\textwidth]{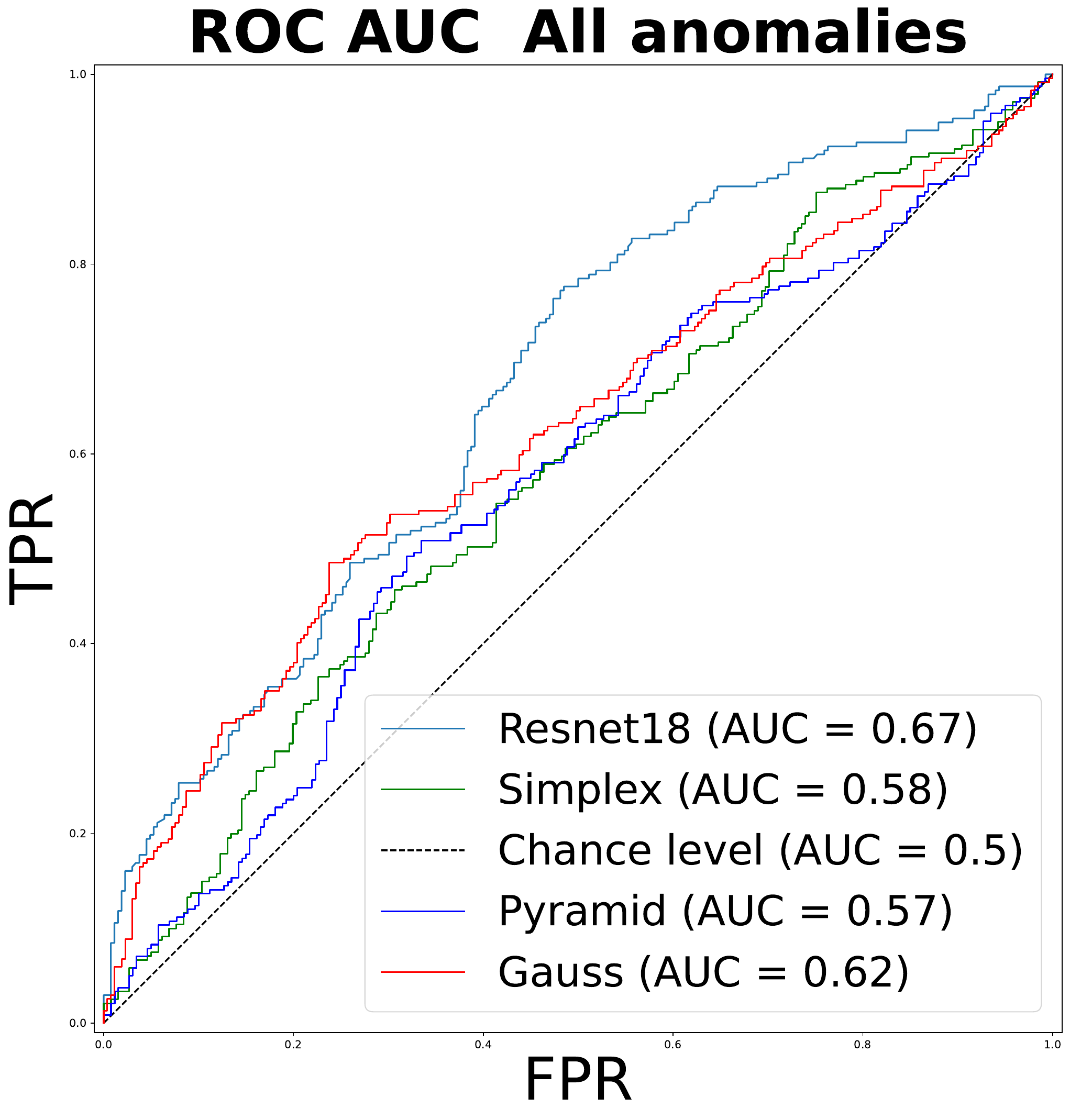}
    \includegraphics[width=0.3\textwidth]{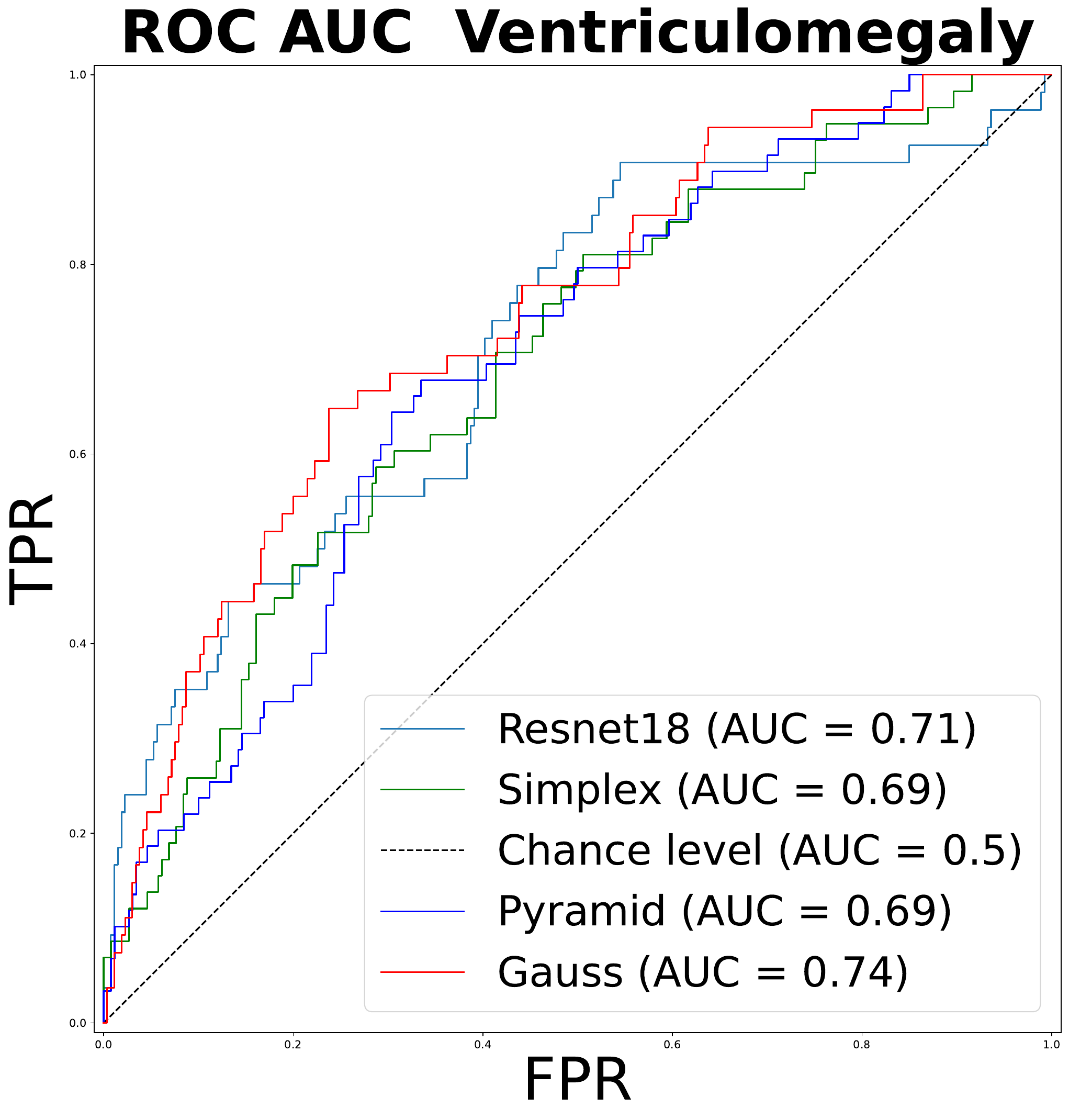}
    \includegraphics[width=0.3\textwidth]{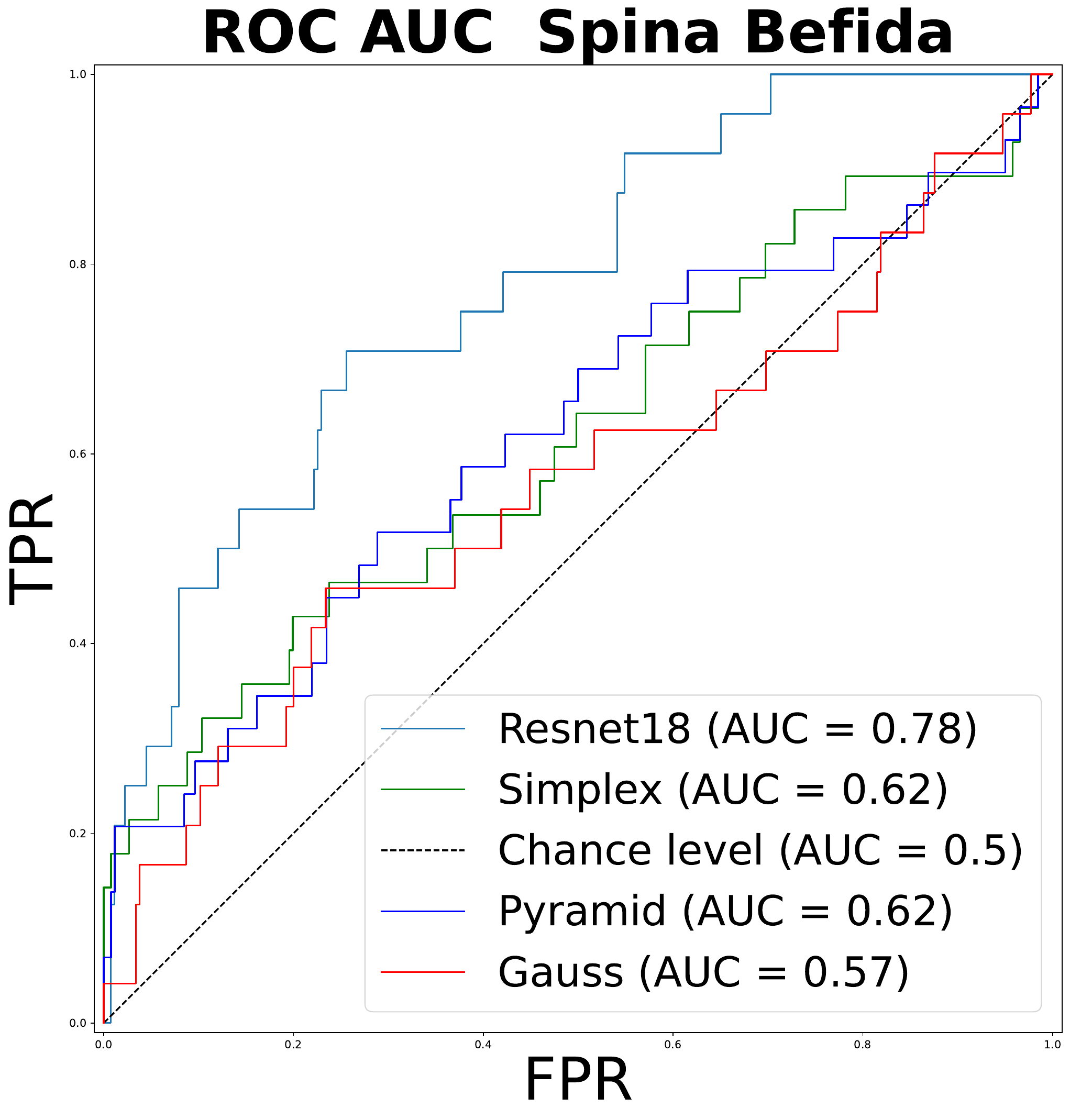}
    \caption{ROC curves on the test set for the different models.
    }
    \label{fig:roccurves}
\end{figure}

We evaluate iNAAD with different noise types for all anomalies and subsets of the most frequent diagnoses, by grouping the infrequent ones into ``Others''. Area Under the Receiver-Operator-Characteristic curve (AUROC) and Average Precision (AP) are reported in Table~\ref{tab:ResultsCodes} and Table~\ref{tab:ResultsCodes2}, respectively. Fig.~\ref{fig:roccurves} illustrates ROC curve examples. The performance of both iNAAD and the supervised vary across different anomaly groups. All iNAAD variants match or exceed the supervised baseline for anomalies that manifest in a localized way, e.g., cerebral cysts, ventriculomegaly, and corpus callosum agenesis. 
For microcephaly, hydrocephalus, and spina bifida, supervised performance is better.

\subsubsection{Ablation study.}
We conducted an ablation study to assess the components of iNAAD. Table~\ref{tab:validation_optimisations} reports AUROC and AP for different similarity metrics and the impact of inpainting and aggregated reconstructions. Note that iNAAD-Gaussian with LPIPS metric, without inpainting, is similar to the method proposed in ~\cite{graham2023denoising}. We observed that the optimal noise level $s$ differs for each noise type $\mathcal{P}$, and pixel-based SSIM outperforms LPIPS and the
semantic similarity metric DeepSim~\cite{czolbe2023semantic} with pre-trained SonoNet-64~\cite{baumgartner2017sononet} as feature extractor. Inpainting the fetal head removes reconstruction errors from anatomically unrelated regions while aggregating reconstruction results in better performance for all noise types. 

\begin{table}[t]
    \centering
    \caption{Ablation study for iNAAD. AUROC and AP results are reported on all anomalies of the validation set (250 ID/250 OOD cases). 
    }
    \scriptsize
    \begin{tabular}{c|c|c|c||c|c}
    \toprule
       $\mathcal{P}$ & $S$ & Similarity Metric & Inpainting & AUROC & AP \\
       \midrule
              & \{150\} & LPIPS &  \ding{55} & 0.54 & 0.53 \\
              & \{150\} & DeepSim & \ding{55} & 0.54 & 0.52 \\
    Gaussian  & \{150\} & SSIM & \ding{55} & 0.58  & 0.59 \\
              & \{150\} & SSIM & \ding{51} & 0.65 & 0.65  \\
              & \{75, 100, 150, 200, 250\} & SSIM & \ding{51} & \textbf{0.68}  & \textbf{0.68} \\
        \midrule
              & \{50\} & LPIPS &  \ding{55} & 0.51 & 0.53 \\
              & \{50\} & DeepSim &  \ding{55} & 0.49 & 0.49 \\
      Simplex & \{50\} & SSIM & \ding{55} & 0.56  & 0.54 \\
              & \{50\} & SSIM & \ding{51} & \textbf{0.59}  & \textbf{0.58}\\
              & \{50, 75, 100\} & SSIM & \ding{51} & 0.58  & \textbf{0.58} \\
       \midrule
                & \{75\} & LPIPS & \ding{55} & 0.55 & 0.54 \\
                & \{75\} & DeepSim &  \ding{55} & 0.52 & 0.50 \\
        Pyramid & \{75\} & SSIM & \ding{55} & 0.57 & 0.55 \\
              & \{75\} & SSIM & \ding{51} & 0.61  & 0.57  \\
              & \{50, 75, 100\} & SSIM & \ding{51} & \textbf{0.62} & \textbf{0.58} \\ \bottomrule
    \end{tabular}
    \label{tab:validation_optimisations}
\end{table}

\subsubsection{Localization and explanability.}
\begin{figure}[h!]
    \centering

    \includegraphics[width=0.49\textwidth,trim={0cm 0cm 0cm 0cm},clip]{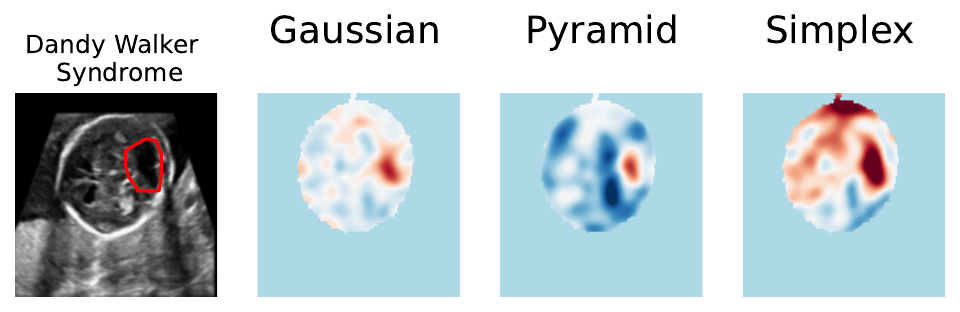}
    \includegraphics[width=0.49\textwidth,trim={0cm 0cm 0cm 0cm},clip]{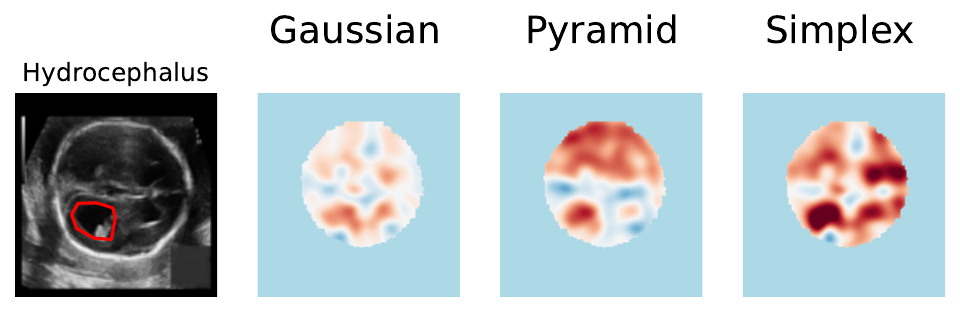}
    
    \includegraphics[width=0.49\textwidth,trim={0cm 0cm 0cm 1cm},clip]{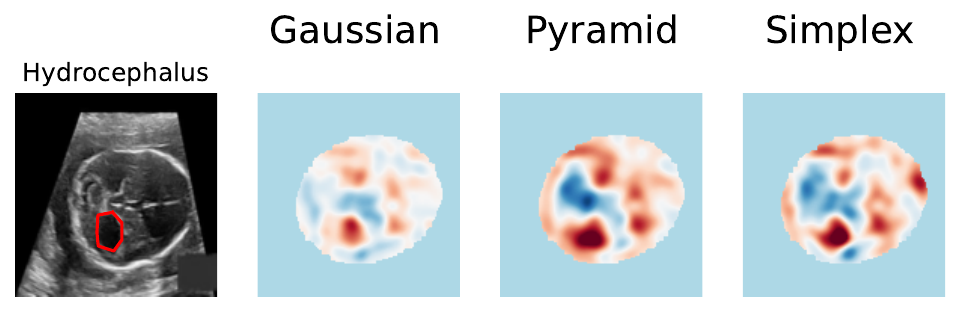}
    \includegraphics[width=0.49\textwidth,trim={0cm 0cm 0cm 1cm},clip]{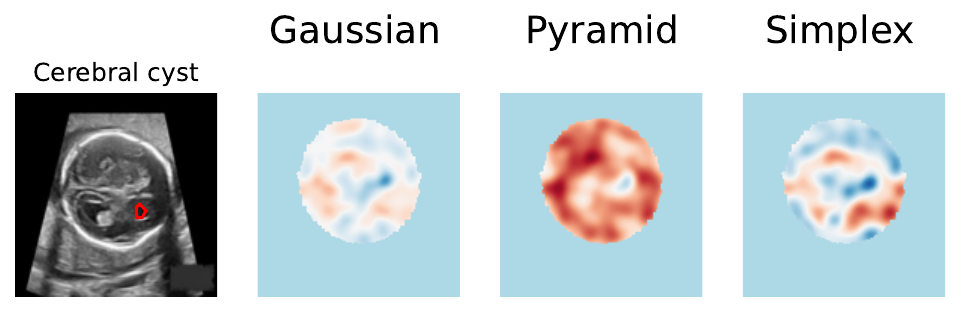}
    
    \includegraphics[width=0.49\textwidth,trim={0cm 0cm 0cm 1cm},clip]{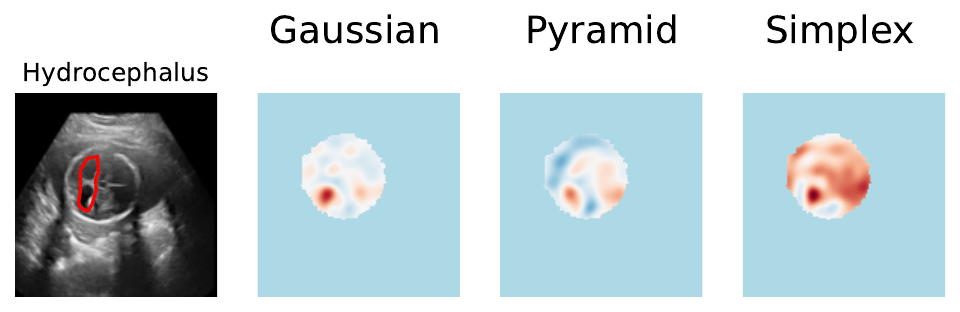}
    \includegraphics[width=0.49\textwidth,trim={0cm 0cm 0cm 1cm},clip]{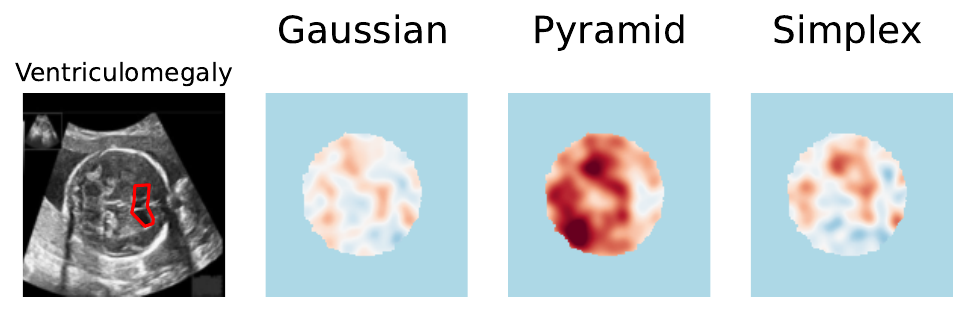}
    
    \includegraphics[width=0.49\textwidth,trim={0cm 0cm 0cm 1cm},clip]{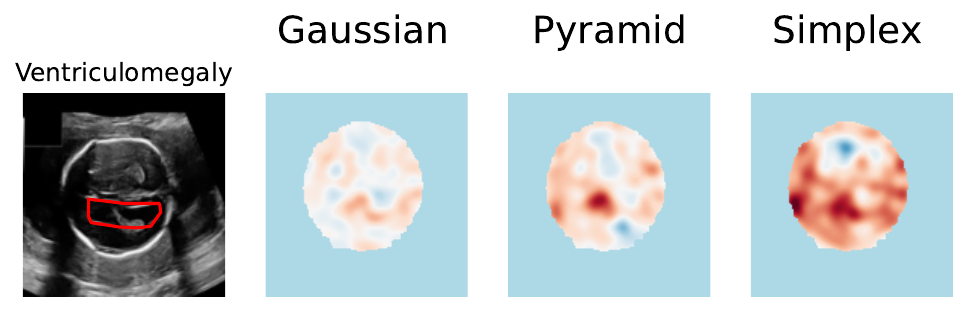}
    \includegraphics[width=0.49\textwidth,trim={0cm 0cm 0cm 1cm},clip]{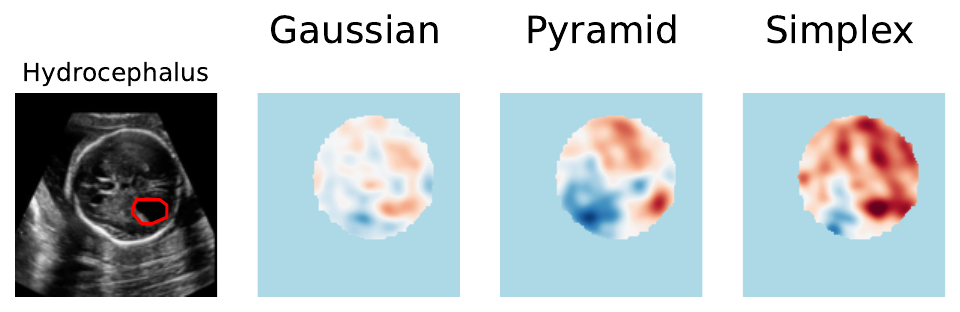}

    \includegraphics[width=0.49\textwidth,trim={0cm 0cm 0cm 1cm},clip]{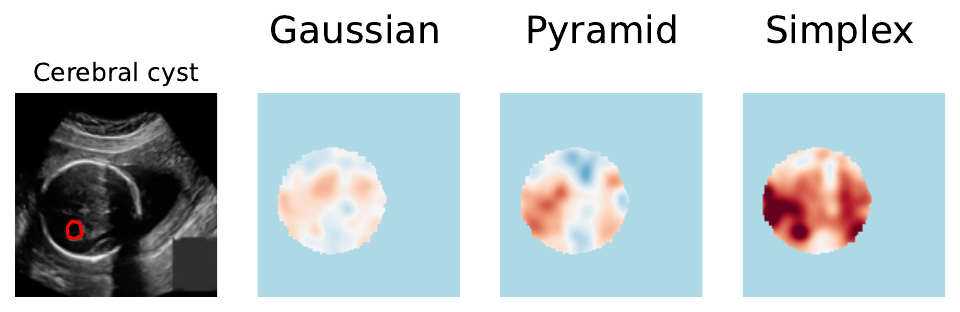}
    \includegraphics[width=0.49\textwidth,trim={0cm 0cm 0cm 1cm},clip]{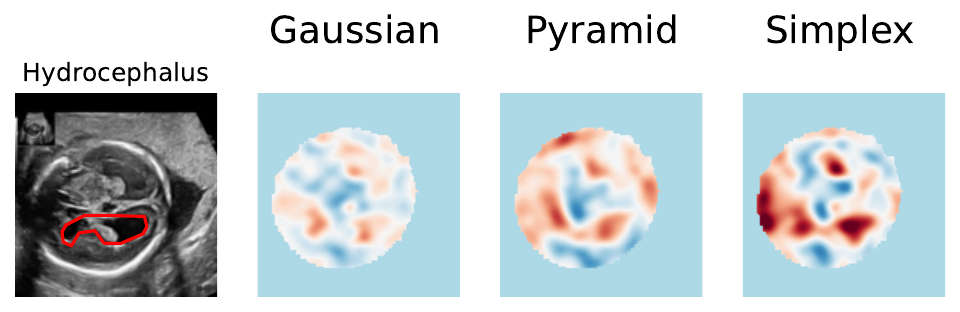}

    \includegraphics[width=0.49\textwidth,trim={0cm 0cm 0cm 1cm},clip]{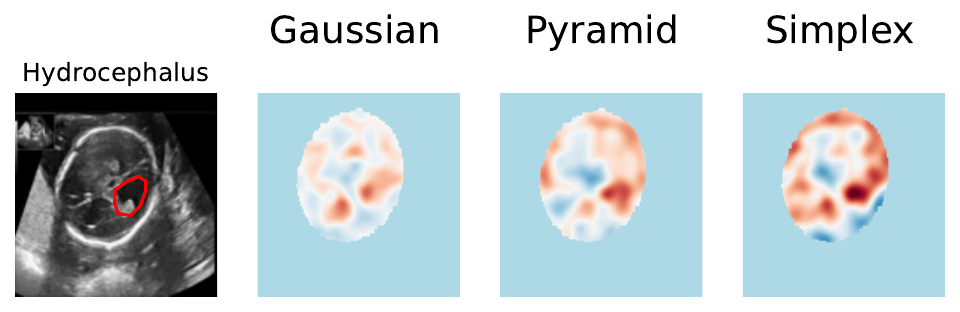}
    \includegraphics[width=0.49\textwidth,trim={0cm 0cm 0cm 1cm},clip]{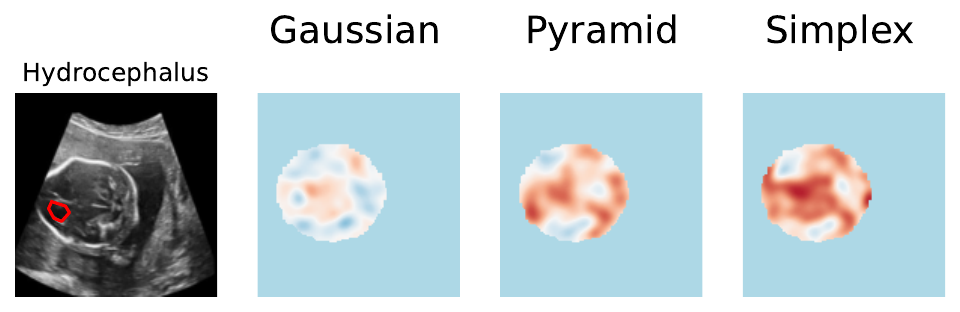}

    \includegraphics[width=0.49\textwidth,trim={0cm 0cm 0cm 1cm},clip]{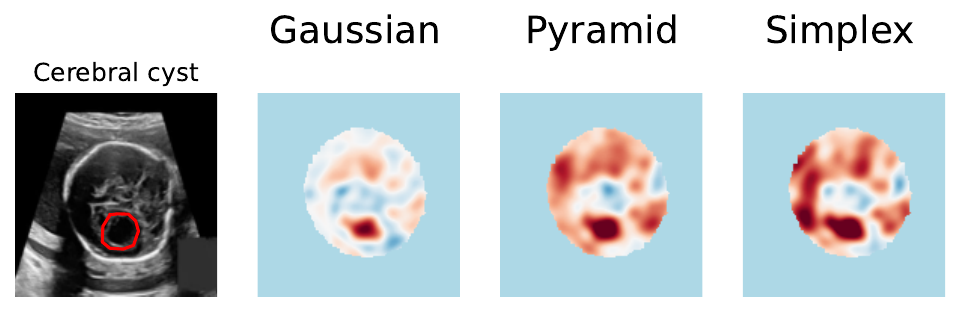}
    \includegraphics[width=0.49\textwidth,trim={0cm 0cm 0cm 1cm},clip]{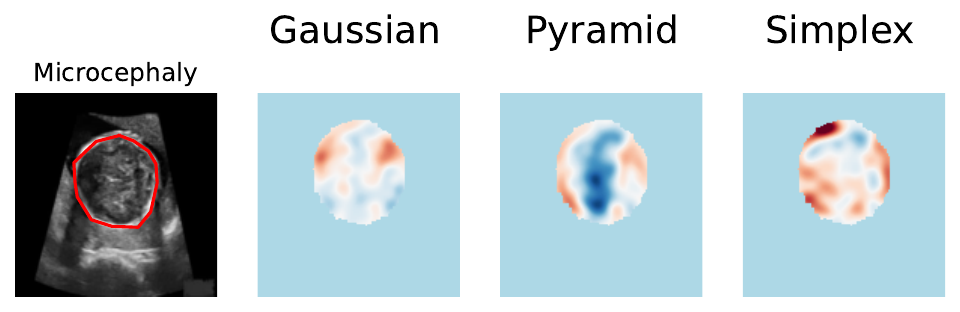}

    \rule{\textwidth}{0.1mm} 
    \hspace{0.01\textwidth}
    
    \includegraphics[width=0.49\textwidth,trim={0cm 0cm 0cm 0.2cm},clip]{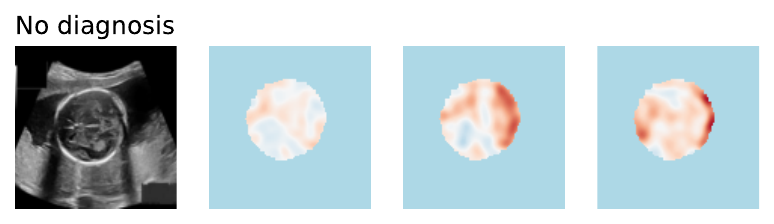}
    \includegraphics[width=0.49\textwidth,trim={0cm 0cm 0cm 0.2cm},clip]{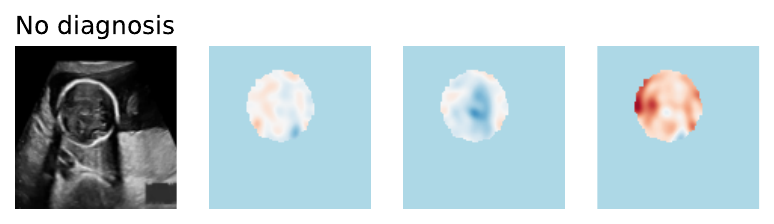}
    \includegraphics[width=0.49\textwidth,trim={0cm 0cm 0cm 1.0cm},clip]{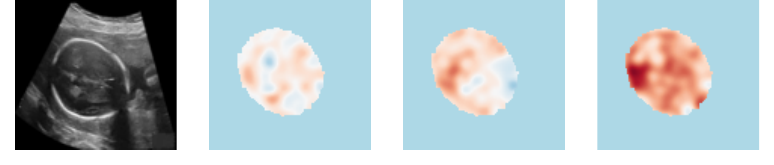}
    \includegraphics[width=0.49\textwidth,trim={0cm 0cm 0cm 1.0cm},clip]{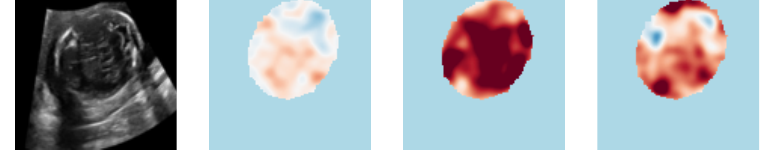}
    \includegraphics[width=0.49\textwidth,trim={0cm 0cm 0cm 1.0cm},clip]{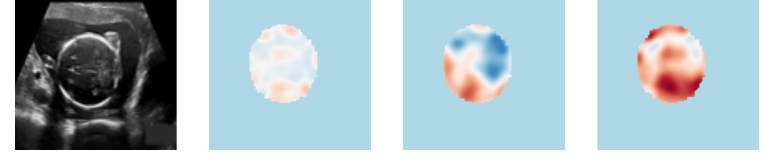}
    \includegraphics[width=0.49\textwidth,trim={0cm 0cm 0cm 1.0cm},clip]{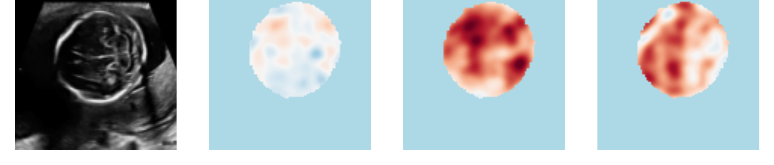}
    
    \caption{Heatmaps and annotated anomalies by an MD with 3 years of experience in prenatal ultrasound imaging. Top: Abnormal cases. Bottom: Normal cases. Anomalies were annotated and localized only for visualization purposes.} 
    \label{fig:qualitative_examples}
    
\end{figure}

 Reconstruction-based methods can be used to segment anomalous regions. Our framework can provide anomaly heatmaps from the reconstruction error offering explainability for localized anomalies such as dandy-walker syndrome, cysts, and hydrocephalus. However, these are less valuable for structural anomalies affecting the entire head, such as microcephaly. Examples of heatmaps for normal and abnormal cases are shown in Fig.~\ref{fig:qualitative_examples}.

\section{Discussion and Conclusion}
Our findings indicate that unsupervised reconstruction-based methods can achieve comparable, in some cases even superior performance, compared to supervised approaches for anomaly detection in medical imaging tasks that are characterized by a scarcity of labeled data for supervised training, but a relative abundance of normal data. Our ablations demonstrate that incorporating inpainting and SSMI as a similarity metric enhances OOD detection of fetal brain anomalies across all noise types.
The proposed method reconstructs normal brains with negligible reconstruction error while providing inherent explainability for localized anomalies as shown in Fig.~\ref{fig:qualitative_examples}. Our experiments on the effect of different noise types show that Gaussian is better on average for the fetal ultrasound setting for image generation, reconstruction, and anomaly detection, unlike MRI settings where Simplex and Pyramid perform best for anomaly detection\cite{wyatt2022anoddpm,frotscher2023unsupervised}. Yet, Simplex noise is better at identifying highly localized anomalies, e.g., cerebral cysts, demonstrating the differences between noise types. Given the low signal-to-noise ratio, anisotropic noise pattern, and orientation-dependence of ultrasound imaging~\cite{asgariandehkordi2023deep}, adapting the noise process for different noise types in fetal ultrasound requires further exploration in future work.

\subsubsection{Limitations.} 
We rely on an automated data extraction process by sampling images from unique patients without manual validation, beyond anatomy identification, to confirm that anomalies are visible in the OOD images. Yet, ensuring non-overlapping patients and diversity in our data splits together with the absence of extensive prepossessing, e.g., including multiple high-quality planes sampled from the same patient videos~\cite{xie2020using}, and removing images with shadows~\cite{xie2020computer}, likely increases the difficulty of our dataset, as reflected by the relatively low performance of our supervised baseline compared to previous studies~\cite{xie2020computer,xie2020using,lin2022use}, whose performance should be interpreted with caution as discussed in~\cite{xie2020using}. Notably, our data reflects real-world conditions, sourced from a national ultrasound screening database, rather than in-depth referral examinations by fetal medicine experts thoroughly examining the brain with the suspicion of an anomaly.
Since previous works rely on extensive annotation, our data may better reflect clinical challenges, emphasizing the need for further clinical validation of all methods.

\subsubsection{Conclusion.} 
We present iNAAD as a proof-of-concept for unsupervised OOD detection using DDPMs to identify fetal brain anomalies. Our approach performs comparably to the supervised baseline on a challenging clinical dataset with a wide range of common fetal brain anomalies, without the need for abnormal cases during training. Finally, iNAAD can serve as a general framework for diffusion-based unsupervised medical anomaly detection with arbitrary noise types and post-hoc adjustments for validation and explainability.

\subsubsection{Acknowledgements.}
This work was supported by the Pioneer Centre for AI (DNRF grant nr P1), the DIREC project EXPLAIN-ME (9142-00001B), the Novo Nordisk Foundation through the Center for Basic Machine Learning Research in Life Science (NNF20OC0062606), and SONAI, an AI signature project from the Danish Agency for Digital Government.

\bibliographystyle{splncs04}
\bibliography{ref}
\end{document}